\documentclass[a4paper,10pt]{article}

\usepackage[ansinew]{inputenc}
\usepackage{amssymb, amsthm}
\usepackage{latexsym}
\usepackage{amsmath}
\usepackage{float, graphicx}
\usepackage[small,bf]{caption}
\usepackage[usenames,dvipsnames]{color}
\usepackage[colorlinks, citecolor=blue, linkcolor=blue, urlcolor=Maroon, filecolor=Maroon]{hyperref}
\usepackage{booktabs}
\makeatother
\usepackage[arrow, matrix, curve]{xy}
\makeatletter
\usepackage{tikz}
\usepackage[abs]{overpic}

\usepackage{geometry}
\allowdisplaybreaks[4]
\theoremstyle{definition}

\newcommand{\sfrac}[2]{{\textstyle\frac{#1}{#2}}}
\newcommand{\und}{\qquad\textrm{and}\qquad}

\newcommand{\ii}{{\rm i\,}}
\newcommand{\dd}{{\rm d}}

\makeatletter
\renewcommand{\@maketitle}{
\newpage
 \null
 \vskip 2em%
 \begin{center}%
  {\Large \textbf \@title \par}%
  \vskip 2.5em
  {\large \@author \par}
  \vskip 1.5 em
 {\large \@date \par}
  \vskip 2.5em 
\end{center}%
 \par}

\title{Instantons, five-branes and fractional strings 
\vspace{0.5cm}}
\author{Christoph N\"olle}
\date{}

\begin{document}
\maketitle
\vspace*{-0.9cm}
\begin{center}
 \em II. Institut f\"ur Theoretische Physik, Universit\"at Hamburg, \\
Luruper Chaussee 149, 22716 Hamburg 
\end{center}

\vspace*{1cm}

\begin{abstract}
 In this note I review the construction of higher-dimensional instantons and heterotic NS5-branes on Ricci-flat cones
from \cite{HN11}, as well as fractional strings from \cite{GHLNP12}. The focus is on methods and
interpretation. I show, furthermore, that smeared 5-brane solutions on manifolds with holonomy $G_2$ or Spin(7) can be
obtained quite generally by a conformal deformation of the Ricci-flat metric, whereas Calabi-Yau and hyperk\"ahler
manifolds require more sophisticated deformations and a case by case treatment.
\end{abstract}

% say mnore explicitly that modified string/gauge-brane suggests 5-brane interpretation! in Sec 3....

% Breite ändern

\numberwithin{equation}{section}

\section{Introduction}
 The NS-sector of heterotic and type II supergravity theories in ten dimensions admits two types of branes: fundamental
strings and their magnetic dual NS5-branes.
Starting from a vacuum solution that contains a 2-dimensional Minkowski space times some Ricci-flat manifold $Y$, the
addition of background strings leads to a warped product of the same geometries, where the inverse of the warp
factor is some harmonic function on $Y$, up to source terms \cite{DGHR90}.
A likewise simple and general construction of NS5-brane backgrounds does not exist, due to the higher dimensionality of
5-branes,
which implies that on a spacetime of the
form $\mathbb R^{9-n,1} \times Y^{n}$ they must wrap cycles $\Sigma \subset Y$ of dimension at least $n-4$, so that
their total world-volume becomes $\mathbb R^{9-n,1} \times \Sigma^{n-4}$. Indeed, several
solutions for 5-branes wrapping calibrated cycles in manifolds of reduced holonomy have
appeared \cite{MaldNun00,AGK00,MaldNast01,GKMW01a,BCZ01,GomisRusso01,GKMW01b,GMPW02}, but all of these constructions are
rather specific to the particular choice of $Y$ and $\Sigma$.
%, and classification results been obtained \cite{GMPW02,GLP05}. 
The original flat space NS5-brane is due to Callan, Harvey and Strominger \cite{CHS91a}. 

In heterotic supergravity the supergravity multiplet couples to a Yang-Mills field, and it has been
argued that Yang-Mills instantons act as a 5-brane source in this theory \cite{Strom90,Witten95,Kanno99}, via the
heterotic Bianchi identity
 \begin{equation}\label{H-Bianchi_intro}
   \dd H = \frac{\alpha'}4 \mbox{tr}\Big(R\wedge R-F\wedge F\Big).
 \end{equation} 
 The moduli space of instantons typically contains a scale parameter $\rho$, which can be
identified with the brane thickness, so for generic values of $\rho$ the brane is smeared, or delocalized. In the
so-called small instanton limit $\rho\rightarrow 0$ the corresponding 5-brane localizes at the position of the
instanton. The prototypical example for this interpretation is given by Strominger's gauge 5-branes \cite{Strom90},
which are based on 4-dimensional BPST instantons and reduce to the localized flat space NS5-branes of
\cite{CHS91a} in the small instanton limit, when the right hand side of \eqref{H-Bianchi_intro} becomes essentially a
delta function supported at the brane locations.

Some further evidence for the interpretation of Yang-Mills instantons in terms of 5-branes came from a result of Tian
\cite{Tian00},
who showed that the singular support of higher-dimensional instantons is always a calibrated subspace of
codimension at least four, which in ten dimensions can be interpreted as the world-volume of a 5-brane. Explicit
examples of instanton based supergravity solutions remained rare, however, and the few existing ones besides
Strominger's gauge 5-branes looked like string or 2-brane backgrounds, rather than 5-branes \cite{HS90,Iva93,GN95};
they are based on the so-called octonionic instantons on $\mathbb R^7$ and $\mathbb R^8$
\cite{FN84,FN85,IP92,IP93,GN95}, whose singular support in the small instanton limit consists of the origin only, so has
codimension strictly larger than
four in ten dimensions. On the other hand, it has been observed in these references that the apparent string or 2-brane
interpretation is problematic as well, since it leads to divergent ADM mass. 

In more recent years, a couple of higher-dimensional instantons on curved spaces has been constructed
\cite{ILPR09,Popov09,HILP10,BILL10,HILP11}. It turned out that particularly useful geometries to
solve the instanton equation are cylinders over certain
homogeneous six- and seven-dimensional manifolds with special SU(3)- and $G_2$-structures, namely nearly
K\"ahler and nearly parallel $G_2$-structures. It is known that the cone over such manifolds admits a holonomy
reduction to $G_2$ and Spin(7), respectively, and due to conformal invariance of the instanton equation, as well as the
fact that the cone metric is conformal to the cylinder metric, the obtained gauge fields also define
instantons on reduced holonomy cones. 

Finally, the attempt of lifting these instantons to heterotic supergravity solutions led to a slight modification of
the instanton construction of \cite{ILPR09,HILP10,BILL10,HILP11}, but also to a generalization to arbitrary
Ricci-flat cones with reduced holonomy (except
odd-dimensional Euclidean spaces in dimension not equal to seven) \cite{HN11}. While the corresponding supergravity
solutions were also found in \cite{HN11}, their interpretation remained elusive at that time, similarly to
the octonionic branes of \cite{HS90,Iva93,GN95} which they contain as special cases. The supergravity metric was found
to be
conformal to the cone metric if the holonomy group of the latter is SU(2), $G_2$ or Spin(7).

The interpretational difficulty with these examples stems from the fact that one cannot straight-forwardly equate the
singular support of the instanton with the brane
world-volume, as Tian's theorem would suggest, because also the corresponding supergravity solution, including the
metric, becomes singular on this hypersurface. This implies that the topology of the hypersurface and hence the
world-volume is not uniquely determined by supergravity. I will argue in Section \ref{ssec:fracstring} that it may
be possible to find an alternative solution which is indeed consistent with the 5-brane interpretation.

\begin{figure}[H]
\begin{center}
\begin{tikzpicture}[domain=0:2]  
 \draw[color=red,thick] plot ({\x},{0.5 + (2/3*( 0.5*\x))^4});            % oben
 \draw[color=red,thick] plot ({\x +2},{0.5 + 16/81 + 64/81*(0.5*\x)});    % oben
 \draw[color=red,thick] plot ({-\x},{0.5});
  \draw[color=red,thick] plot ({\x},{-0.5 - (2/3*( 0.5*\x))^4});            % unten
 \draw[color=red,thick] plot ({\x +2},{-0.5 - 16/81 - 64/81*(0.5*\x)});    % unten
  \draw[color=red,thick] plot ({-\x},{-0.5});
 \draw[color=black] plot ({4 - 0.3*sin((\x r)*pi/2)},{(0.5+80/81)*cos((\x r)*pi/2)}); % Kreis 1
 \draw[color=black,dashed] plot ({4 + 0.3*sin((\x r)*pi/2)},{(0.5+80/81)*cos((\x r)*pi/2)}); % Kreis 1
  \draw[color=black] plot ({2 - 0.2*sin((\x r)*pi/2)},{(0.5+16/81)*cos((\x r)*pi/2)}); % Kreis 2
 \draw[color=black,dashed] plot ({2 + 0.2*sin((\x r)*pi/2)},{(0.5+16/81)*cos((\x r)*pi/2)}); % Kreis 2
  \draw[color=black] plot ({ - 0.15*sin((\x r)*pi/2)},{(0.5)*cos((\x r)*pi/2)}); % Kreis 3
 \draw[color=black,dashed] plot ({ 0.15*sin((\x r)*pi/2)},{(0.5)*cos((\x r)*pi/2)}); % Kreis 3
  \draw[color=black] plot ({ -2- 0.15*sin((\x r)*pi/2)},{(0.5)*cos((\x r)*pi/2)}); % Kreis 4
 \draw[color=black,dashed] plot ({ -2+0.15*sin((\x r)*pi/2)},{(0.5)*cos((\x r)*pi/2)}); % Kreis 4
\end{tikzpicture} \caption{NS5-brane geometry. The left end is an infinite cylinder over a base manifold $X$, carrying
a linear dilaton supergravity solution, and the right end is a Ricci-flat cone over $X$ with a vacuum solution. Black
circles represent the base $X$. The worldvolume of the branes is located at an
infinite geodesic distance to the left, and the asymptotic radius of the cylinder is proportional to
the number of branes.}\label{fig:NS5brane}
\end{center}
\end{figure}
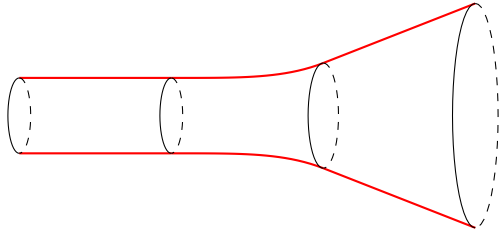

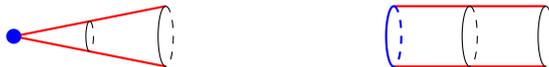
\begin{figure}[H]
\begin{center}
\begin{tikzpicture}[domain=0:2]  
 \draw[color=red,thick] plot ({\x},{0.2*\x});  % Cone
 \draw[color=red,thick] plot ({\x},{-0.2*\x}); % Cone
 \draw[color=red,thick] plot ({\x+5},{0.4});   % Cylinder
 \draw[color=red,thick] plot ({\x+5},{-0.4});   % Cylinder 
 \draw[color=black] plot ({7 - 0.1*sin((\x r)*pi/2)},{(0.4)*cos((\x r)*pi/2)}); % Kreis 1
 \draw[color=black,dashed] plot ({7 + 0.1*sin((\x r)*pi/2)},{(0.4)*cos((\x r)*pi/2)}); % Kreis 1
 \draw[color=black] plot ({6 - 0.1*sin((\x r)*pi/2)},{(0.4)*cos((\x r)*pi/2)}); % Kreis 2
 \draw[color=black,dashed] plot ({6 + 0.1*sin((\x r)*pi/2)},{(0.4)*cos((\x r)*pi/2)}); % Kreis 2
 \draw[color=blue,thick] plot ({5 - 0.1*sin((\x r)*pi/2)},{(0.4)*cos((\x r)*pi/2)}); % Kreis 3
 \draw[color=blue,dashed,thick] plot ({5 + 0.1*sin((\x r)*pi/2)},{(0.4)*cos((\x r)*pi/2)}); % Kreis 3
 \draw[color=black] plot ({2 - 0.1*sin((\x r)*pi/2)},{(0.4)*cos((\x r)*pi/2)}); % Kreis 4
 \draw[color=black,dashed] plot ({2 + 0.1*sin((\x r)*pi/2)},{(0.4)*cos((\x r)*pi/2)}); % Kreis 4
 \draw[color=black] plot ({1 - 0.05*sin((\x r)*pi/2)},{(0.2)*cos((\x r)*pi/2)}); % Kreis 5
 \draw[color=black,dashed] plot ({1 + 0.05*sin((\x r)*pi/2)},{(0.2)*cos((\x r)*pi/2)}); % Kreis 5
 \node[circle,scale=0.6,fill=blue]{};
\end{tikzpicture} \caption{Two partial compactifications of an open cylinder, giving rise to a cone and a cylinder with
boundary. The blue parts on the left boundary are the added pieces, consisting of a single point in the left picture,
and a copy of the cylinder base in the right one. Since the dimension of the added boundaries differs, the
world-volume of a brane located at the boundary can be higher in the right picture. While the
correct dimension of the gauge branes cannot be deduced from the corresponding supergravity solutions, the coupled
string - gauge brane solutions indicate that the compactification of the right picture is the physical one. For the
geometries relevant to supergravity it is spacious enough to accommodate a 5-brane, contrary to
the one point compactification.}\label{fig:cylcptfc}
\end{center}
\end{figure}

To illustrate this phenomenon consider the metric of the gauge brane solutions of \cite{HN11} in the small instanton
limit:
 \begin{equation}\label{intro:5-branemetric}
   g = \eta_{\mu\nu}\dd x^\mu \dd x^\nu + \Big(1 + \frac {Q_m}{r^2} \Big) \big( \dd r^2 + r^2g^k\big),
 \end{equation} 
 where $g^k$ is the metric on the $k$-dimensional base $X^k$ of the cone, and $Q_m$ is a brane charge proportional to
the number of branes. In the asymptotic region $r\rightarrow \infty$ the metric
\eqref{intro:5-branemetric} approaches the cone metric, which is simply the Euclidean metric on $\mathbb R^{k+1}$ in
the case $X=S^k$, and in the near horizon limit $r\rightarrow 0$ one obtains a
cylinder metric, as illustrated in Figure \ref{fig:NS5brane}. 
The singular support of the gauge field in the small instanton limit, and hence the world-volume
of the brane, is located at the infinitely remote boundary of the cylinder at $r=0$, which is
not part of physical spacetime, since light-like geodesics cannot reach this region in finite time, and neither
can probe strings.  
Since the boundary at $r=0$ is not part of physical spacetime, its topology is not determined by the supergravity
equations, and there are indeed several possible (partial) compactifications of the open cylinder. A common choice is
the one-point compactification, leading to a conical topology, whereas addition of a copy of the base $X$ gives rise to
an open cylinder with boundary, see Figure \ref{fig:cylcptfc}.
Of course, the dimension of the boundary plays an important role in the interpretation of the solution, since it
contains the world-volume of the brane. The one-point compactification of the cylinders over $S^6$ and $S^7$
could at most give rise to string or 2-brane world-volumes, which is what confused the authors of
\cite{HS90,Iva93,GN95}.
The cylinder compactification on the other hand admits a 5-brane world-volume, which would have to wrap
certain codimension three cycles inside the base $X$.

For the octonionic instantons this means that one has to replace the origin of $\mathbb R^7$ or $\mathbb R^8$ by a
sphere of dimension six or seven, and the 5-brane is expected to wrap appropriate smeared
cycles of dimension three or four inside this sphere. The nature of these wrapped cycles is not well understood yet,
except for the flat space NS5-brane with $X=S^3$, where they can be viewed as a superposition of all points of $S^3$,
and the brane is smeared evenly over the three-sphere. In general, the cycles should be calibrated with respect to a
canonical $(k-3)$-form $*P$ on the base manifold, according to Tian's theorem.

By superposing 5-branes with fundamental strings another class of heterotic supergravity solutions can be obtained.
Suppose that $Y^n$ ($n\leq 8$) is a Ricci-flat manifold with reduced holonomy, $\Sigma^{n-4}\subset Y^n$ a calibrated
submanifold and $\mathbb T^{8-n}$ a flat torus. An NS5-brane wrapping $\Sigma^{n-4}\times \mathbb T^{8-n}$
inside the space-time $\mathbb R^{1,1}\times Y^n\times \mathbb T^{8-n}$ looks effectively like a string, since its
world-volume has two non-compact directions. By adding additional fundamental strings to this background, one ends up
with fractional strings, consisting of both genuine and `fake' strings. Explicit solutions of this type
have been constructed on cones with reduced holonomy in \cite{GHLNP12},\footnote{fractional strings are sometimes
called ``dyonic strings'' \cite{DFKR95}, and the term ``NS1-NS5-brane'' is used in \cite{GHLNP12}. Since they are
completely analogous to fractional branes, the name ``fractional strings'' appears most appropriate to me now.} based on
the
smeared 5-branes of \cite{HN11}. Their near horizon geometry is of the form AdS$_3\times X^{n-1}\times
\mathbb T^{8-n}$, with $X^{n-1}$ an Einstein manifold and $n\in\{4,6,7,8\}$, which makes them promising candidates for
AdS$_3$/CFT$_2$-duality.

These solutions might also provide some further evidence for the 5-brane interpretation of gauge branes. Again, the
world-volume of the strings and branes is not part of the physical spacetime, due to a metric singularity. But this time
the singularity can be removed by hand, to obtain another solution which is completely smooth and extends to negative
values of the radial variable $r$.
 One might then try to determine the singular support of an instanton on this new geometry. There are,
however, two issues with the new solution. Firstly, due to a coordinate singularity at $r=0$ it is not straight-forward
to determine
the continuation explicitly, and in fact it has not been found so far. An educated guess would be a solution which is
reflection symmetric in $r$, giving rise to a wormhole geometry as in Figure \ref{fig:wormhole} below. This is known
to solve the equations for $r\neq0$, and is continuous at $r=0$, so only smoothness at $r=0$ remains an open question.
The second issue is that the removal of the metric singularity also eliminates the singularity of the instanton
that occurs in the supergravity solution, which was our main object of interest. This new instanton does not define a
gauge field
on the cone, but is given simply by the pull-back of a so-called canonical connection $\nabla^P$ on $X$ to the full
regular geometry, which topologically is a direct product containing $X$. Away from $r=0$, the small instanton on the
cone coincides with the pull-back of $\nabla^P$.

Nevertheless, once the full solution is known it must be possible to extend the cone
instantons defined in the region $r>0$ to negative $r$ values as well, and determine the singular support in the small
instanton limit. 
Assuming again reflection symmetry in $r$, the singular support would have to be contained in the
hypersurface $\{r=0\}$, which is a submanifold of
AdS$_3\times X^{n-1}\times \mathbb T^{8-n}$ defined by the vanishing of the radial Poincar\'e coordinate of AdS$_3$.
Due to the increased dimension of the hypersurface in the new solution as opposed to the cone, we can expext that the
dimension of the singular support gets blown up to its maximum allowed value six as well.
It should be emphasized that while all this is fully
compatible with the 5-brane interpretation we advocated above in terms of a cylinder compactification with boundary, 
there does not appear to be any compelling reason why the singular support of a small instanton on the modified
string-gauge brane geometry should determine the brane world-volume, and not for instance its singular support on the
Ricci-flat cone, which is also smooth in the case $X=S^k$.

 Note also that this
program requires a higher-dimensional interpretation of the original instantons on the cone over $X$. For
instance, one would have to extend the 4-dimensional BPST instantons on $\mathbb R^4$ to gauge fields on a
6-dimensional Lorentzian manifold in order to determine the world-volume of the ordinary flat space NS5-brane in this
way.

The simple construction of 5-branes in terms of conformal transformations of reduced holonomy cone metrics applies only
to the case of SU(2), $G_2$ or Spin(7) holonomy on either four-, seven- or eight-manifolds, whereas the holonomy groups
SU($m$) and Sp($m-1)$ for $m>2$ require a somewhat more sophisticated approach. For simplicity I will concentrate
on the first class of holonomy groups in this review, which can be treated in a unified way. It turns out that the
conical nature of the metric is not essential for most of the construction; the gravitino and dilatino BPS equations can
be solved for arbitrary conformal transformations of the reduced holonomy metric, and it is only the gauge sector where
we need the conical structure to construct explicit BPST-like instantons, which then determine the conformal factor via
the Bianchi identity \eqref{H-Bianchi_intro}. It appears likely that more 5-branes can be constructed with this
approach, using explicit instanton gauge fields on other manifolds with exceptional holonomy group, for instance those
based on generalized Killing spinors \cite{Baer_generalizedcylinder,ChiossiSal02,ContiSal05,CLSS09}.

 The article starts in Section \ref{sec:generalRicciflat} with a discussion of heterotic supergravity
on manifolds with reduced holonomy, with particular focus on string and 5-brane solutions. In Section \ref{sec:cones} we
specify to cones and discuss explicit solutions, again for strings and 5-branes, and finally the superposition of
both.

\section{Strings and 5-branes on Ricci-flat manifolds}\label{sec:generalRicciflat}
 \subsection{Heterotic supergravity}
 The NS sector of IIA, IIB and heterotic supergravities has bosonic field content $(g,H,\phi )$, where $g$ is a
Lorentzian metric, $H$ a 3-form and $\phi$ a function, the dilaton. In general they have to satisfy a set of second
order field equations, and for a background to preserve supersymmetry the following first-order BPS equations must be
imposed:
 \begin{equation}\label{BPSeqtns}
\begin{aligned}{}
    \Big( \nabla_\mu - \frac 18 H_{\mu\nu\lambda} \gamma^{\nu\lambda}\Big) \epsilon &= 0 ,\\
      \big(\dd \phi -\sfrac 12 H\big) \cdot \epsilon &=0,
\end{aligned}
 \end{equation} 
 where $\epsilon$ is a Majorana-Weyl spinor, $\nabla$ the Levi-Civita connection for $g$, and the Clifford action of
a form on a spinor is denoted by a dot. Additionally, the 3-form has to satisfy a Bianchi identity, which is $\dd H=0$
at lowest order in $\alpha'$. The BPS equations do not receive stringy corrections at order $(\alpha')^1$, but at least
in the heterotic theory the Bianchi identity does:
 \begin{equation}\label{H-Bianchi}
   \dd H= \frac {\alpha'}4 \mbox{tr}\Big(R\wedge R -F\wedge F\Big),
 \end{equation} 
 where $R$ and $F$ are both Yang-Mills field strengths, i.e. Lie-algebra valued
2-forms. The curvature form $R$ is associated to a connection $\tilde \nabla$ on the tangent bundle, and $F$ to a
connection $\nabla^A$ on some arbitrary vector bundle, whose
structure group should be contained in $E_8\times E_8$ or SO(32) for string theory
applications. There is a BPS equation for $F$ as well, namely the instanton equation
 \begin{equation}
   F\cdot \epsilon=0,
 \end{equation} 
 and it has been shown that if one imposes the same condition on $R$, i.e. $R\cdot\epsilon=0$, then under certain mild
conditions the BPS equations and $H$-Bianchi identity imply the field
equations \cite{Iv09}. These conditions are satisfied for pure 5-brane solutions, but for string configurations
there is one additional independent field equation that has to be imposed, the $H$-equation 
 \begin{equation}\label{H-eqtn}
   \dd *(e^{-2\phi } H) =0.
 \end{equation} 

\subsection{Ricci-flat solutions}
 A simple solution to the BPS equations is obtained by setting $H=\dd \phi =0$ and using a direct product metric on
$\mathbb R^{9-k,1} \times Y^n$. The gravitino equation reduces to 
 \begin{equation}
   \nabla \epsilon =0 ,
 \end{equation} 
 so $Y$ must admit a parallel spinor. This implies a reduction of the holonomy group of $Y$ to a subgroup $G\subset
\mbox{SO}(n)$ which leaves a spinor invariant. The possible irreducible holonomy groups have been classified by Wang:
  \renewcommand{\arraystretch}{1.2}
  \begin{table}[H]\centering
 \begin{tabular}{ccccc}  \toprule
      $n$ & $G$ &  $\sharp$ spinors & & geometry\\ \hline 
        7  & $G_2 $ & 1 &  &exceptional\\
        8  & Spin(7) & (1,0) & & exceptional\\ 
        $2m$ & SU($m$) & (1,1) & ($m$ odd) & Calabi-Yau\\
        $2m$ & SU($m$) & (2,0) & ($m$ even)& Calabi-Yau\\
        $4m$ & Sp($m$) & ($m+1$,0) & &hyperk\"ahler\\       
 \bottomrule
\end{tabular}
 \caption{Invariant spinors for subgroups $G\subset \mbox{SO}(n)$. In even dimensions even and odd chirality
spinors are listed separately. Subgroups of $G$ can fix additional spinors and lead to enhanced supersymmetry.}\label{} 
\end{table}
 It is known that every metric with one of the above holonomy groups is Ricci-flat, which is also required by the field
equations.
 The $H$-Bianchi identity as well as the instanton equations for $F$ and $R$ are solved if we choose $F$ and $R$ equal
to the Riemannian curvature form associated to the Ricci-flat metric $g$.

 It is very difficult to find explicit metrics with the above holonomy groups, in fact no analytical solutions are
known on compact spaces, although existence has been established in all cases. There are, however, some rather simple
non-compact examples, in particular cones over certain Einstein manifolds. These will be
discussed in Section \ref{sec:cones}.

\subsection{Harmonic deformations: strings}\label{subsec:harmdef}
 Consider a spacetime of the form $\mathbb R^{1,1} \times Y^n \times \mathbb T^{8-n}$, where $(Y^n,g_0)$ is Ricci-flat
and $\mathbb T^{8-n}$ is a torus, equipped with the following fields:
 \begin{equation}\label{strings:fields_general}
  \begin{aligned}
     g &= h^{-1} (-\dd t^2+\dd x^2) + g_0 + g_{\mathbb T^{8-n}}, \\
     H &= \dd h^{-1} \wedge \dd t \wedge \dd x,\\
     e^{2(\phi-\phi_0)} &= h^{-1},
  \end{aligned}
 \end{equation} 
 with $h$ a harmonic function on $Y^n$, up to possible delta function singularities. This solves the field equations, up
to possible source terms, and also the BPS
equations if $Y$ admits a parallel spinor. The interpretation of the solution is in terms of a number of strings placed
on the Ricci-flat background $\mathbb R^{1,1} \times Y^n\times \mathbb T^{8-n}$. The function $h$ must satisfy a
quantization condition, which gives rise to a discrete parameter that can be identified with the number of strings.
Since the 3-form $H$ is closed for this background it is not necessary to consider the $\alpha'$-corrected Bianchi
identity, although it can be solved easily by setting $\tilde \nabla$ and $\nabla^A$ equal to the Levi-Civita connection
of the Ricci-flat background.
% Note that for $Y$ to admit a non-trivial harmonic function it must be non-compact.
% what if source terms are present?

\subsection{Conformal deformations: 5-branes}\label{subsec:confdef}
 Besides the simple Ricci-flat solutions and their harmonic deformations above, metrics with reduced holonomy can be
used to construct additional solutions with non-vanishing fluxes. Suppose that $(Y^n,g_0)$ has a parallel spinor, then
the reduced holonomy group $G$ gives rise to a parallel 4-form $Q$ on $Y$, which is related to the Casimir element of
$G$. On a $G_2$ manifold with defining 3-form $\Phi$ we have $Q=*\Phi$, on a Spin(7)-manifold $Q$ is equal to the
Cayley-form $\Psi$, and on a Calabi-Yau $Q = \frac 12 \omega\wedge \omega,$ where $\omega$ is the K\"ahler form.
Consider the following fields on $Y$:
 \begin{equation}\label{5branes:fields_general}
\begin{aligned}{}
    g &= e^{2f} g_0, \\  
   H& = - 2\beta^{-1}e^{4f}\,\dd f \lrcorner\, Q ,\\ 
   \phi &= \phi_0 +  \sfrac 13(n-1)f  ,
\end{aligned}   
\end{equation} 
 where $f$ is any function on $Y$, $\phi_0$ is the asymptotic value of the dilaton, and $\beta $ is a 
constant parameter, to be determined shortly.
The contraction of forms $\dd f\lrcorner\, Q$ is taken with respect to the conformally rescaled metric. Implicitly  it
is
understood that the spacetime and metric contain another flat Minkowski space factor, such that the total dimension
adds up to ten. The new metric $g$ no longer admits a parallel spinor, but it turns out that if $G$ is either
SU(2), $G_2$ or Spin(7), then the above ansatz still gives rise to a solution of the gravitino and dilatino equations
\eqref{BPSeqtns} for
appropriate choice of $\beta$, due to the following relation (stated for the
Ricci-flat metric):\footnote{\eqref{Q-epsilon-action} and
\eqref{Q-epsilon-action2} are most conveniently proven on a case by case basis, using a standard form for $Q$ and
$\epsilon$. Alternatively, one can express $Q$ as a spinor
bilinear in $\epsilon$ and use Fierz identities.}
\begin{equation}\label{Q-epsilon-action}
 \frac 12 Q _{abcd}\gamma^{cd} \epsilon = \beta\, \gamma_{ab}\epsilon ,
\end{equation}
 where $\epsilon$ is a $G$-invariant spinor, and
  \begin{equation}\label{beta}
   \beta = \begin{cases}
               1 & \text{for SU(2)},\\
               2 & \text{for }G_2,\\
	       3 & \text{for Spin(7)}. \\
            \end{cases}
 \end{equation} 
Our choice of fields together with \eqref{Q-epsilon-action} guarantees that the $H$-term in the gravitino equation $(
\nabla_a - \frac 18H_{abc}\gamma^{bc})\epsilon=0$ exactly compensates the correction to the
Levi-Civita connection induced by the conformal rescaling of the metric. Furthermore, the dilatino equation $(\dd \phi
-\frac 12 H)\cdot \epsilon=0$ is satisfied because of
 \begin{equation}\label{Q-epsilon-action2}
  \frac 16Q_{abcd} \gamma ^{bcd} \epsilon = -\kappa\gamma_a \epsilon,
 \end{equation} 
 with
  \begin{equation}\label{kappa}
   \kappa = \begin{cases}
            \ \    4 &\text{for }G_2,\\
	     \ \  7 & \text{for Spin(7)}, \\
		m-1 & \text{for SU($m$)},\\
                \frac{2m+3}3     &  \text{for Sp($m$)}.\\
            \end{cases}
 \end{equation} 
 Note that \eqref{Q-epsilon-action} implies \eqref{Q-epsilon-action2} with 
 \begin{equation}
   \kappa = \frac \beta 3 \big(n-1\big) .
 \end{equation} 
 For $G=\mbox{SU}(m)$ or Sp($m-1)$ with $m>2$ the Clifford action of
$Q_{\mu\nu\rho\sigma}\gamma^{\rho\sigma}$ on
$\epsilon$ is more complicated than in \eqref{Q-epsilon-action}, e.g. for SU($m$) we have
 \begin{equation}
   \frac 12 Q_{abcd}\gamma^{cd} \epsilon = \big(\gamma_{ab} -\ii (m-2) \omega_{ab} \big)
\epsilon,
 \end{equation}
and I am not aware of a general procedure to solve the
BPS equations in these cases. In the specific example of conical Ricci-flat manifolds, however, it is possible to obtain
very similar solutions for these holonomy groups by deforming the metric in a slightly more complicated way \cite{HN11}.

The space-like 3-form flux of the solution \eqref{5branes:fields_general} to the BPS equations implies that it describes
the back-reacted geometry of an NS5-brane wrapping some cycles in the reduced holonomy manifold $Y$. Generically, the
3-form $H$ is not closed, so the zeroth order Bianchi identity does not hold. There is however the possibility to
include the first
heterotic $\alpha'$ correction and impose equation \eqref{H-Bianchi}, with a 5-brane source term leading to an
interpretation in terms of smeared 5-branes. To complete the solution we need to find two
instanton connections $\nabla^A$ and $\tilde \nabla$ whose curvature forms $F$ and $R$ satisfy
 \begin{equation}
   \frac{\alpha'}4 \mbox{tr} \Big(R\wedge R -F\wedge F \Big ) = \dd H.
 \end{equation} 
 As we have already observed, there is a canonical instanton connection on $Y$, the Levi-Civita connection $\nabla^0$ of
the Ricci-flat metric. Typically, instantons possess moduli and it may be possible to choose both connections as
deformations of $\nabla^0$. For fixed choice of connections the Bianchi identity should
determine the function $f$ completely, up to an overall volume parameter. This is indeed the case if $Y$ is a cone;
Strominger's gauge 5-branes \cite{Strom90} for
instance are obtained by taking $Y= \mathbb R^4$, the cone over $S^3$, and setting $\tilde \nabla$ equal to the flat
connection $\nabla^0$, whereas $\nabla^A$ is chosen as a BPST instanton. The relevant group in this case is
$G=\mbox{SU}(2)$, although the holonomy group of $\mathbb R^4$ is trivial. The moduli space of a single BPST
instanton contains a scale parameter $0\leq\rho<\infty$. For generic $\rho$ one
obtains a gauge 5-brane,
the small instanton limit $\rho\rightarrow 0$ gives the NS5-brane \cite{CHS91a}, and for $\rho\rightarrow \infty$ the
instanton and brane disappear off to infinity, leaving us with the empty Minkowski space solution, with flat
connections.
We will see below that the picture remains roughly the same when we replace $\mathbb R^4$ by a higher-dimensional cone.
A specific property of the SU(2)-case is that the right hand side of the Bianchi identity vanishes in the small
instanton limit, except for a delta-function singularity at the location of the instanton. Hence, the brane is fully
localized in this case.
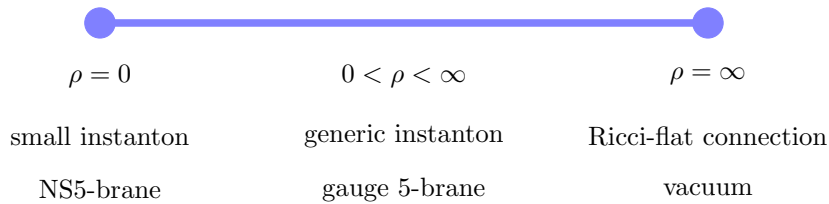
\begin{figure}[H]
\begin{center}
\begin{tikzpicture}
   \draw[color=blue!50,line width=1mm](-4,0) -- (4,0);
   \node[draw,circle,fill=blue!50,color=blue!50,minimum height=0.4cm] at (-4,0) (null) {};
     \node[draw,circle,fill=blue!50,color=blue!50,minimum height=0.4cm] at (4,0) (infty) {};
  \node at (-4,-0.7) {$\rho=0$}; 
  \node at (0,-0.7) {$0< \rho<\infty$};
  \node at (4,-0.7) {$\rho=\infty$};
    \node at (-4,-1.5) {small instanton};
   \node at (0,-1.5) {generic instanton};
  \node at (4,-1.5) {Ricci-flat connection};
  \node at (-4,-2.2) {NS5-brane};
     \node at (0,-2.2) {gauge 5-brane};
  \node at (4,-2.2) {vacuum};
% \node[below=0.8] at (-1,-1.732) {\footnotesize $\exp(-2\pi\, \ii/3)$};
%\node[above=0.8] at (-1,1.732) {\footnotesize $\exp(2\pi\, \ii/3)$};
% \node at (2.2,0) {\footnotesize $1$};
%  \node at (0.1,-0.2) {\footnotesize $0$};
\end{tikzpicture} \caption{The scale parameter $\rho$ of a single instanton on a Ricci-flat cone and the corresponding
supergravity solutions. Here it is assumed that $\tilde \nabla$ coincides with the Ricci-flat
connection, and $\nabla^A$ is an instanton characterized by the size $\rho$. Gauge 5-branes are delocalized or smeared
NS5-branes.}\label{fig:Inst_scale}
\end{center}
\end{figure}
Solutions to the gravitino and dilatino equations have been classified in terms of
$G$-structures \cite{Ivanov01,FI01b,GMPW02,GLP05}, and the conformally Ricci-flat solutions presented in this section
are included in these results, although they do not receive much attention in the references. It is found for instance
in \cite{FI01b} that any conformally balanced $G_2$-manifold in dimension seven leads to a solution, and likewise in
\cite{Ivanov01} for 8-dimensional manifolds with conformally balanced Spin(7)-structure. Conformally balanced means that
the Lee 1-form
 \begin{equation}
   \theta =  Q \lrcorner\, \dd  Q 
 \end{equation} 
 is exact; it will be proportional to $\dd\phi$ in the supergravity solution. In particular, a conformally
rescaled metric of holonomy $G_2$ or Spin(7) belongs to this class. 

Conformally Ricci-flat solutions also appeared in the context of orientifold compactifications in type II
supergravity \cite{GKP01,BDJRWZ10}, with additional RR-form fluxes. In these examples smeared source terms lead to
Ricci-flat internal spaces, whereas the conformal deformation occurs only for localized orientifolds \cite{BDJRWZ10}.

\section{Strings, instantons and 5-branes on a cone}\label{sec:cones}
 The simplest non-trivial spin manifolds with parallel spinors are cones over certain Einstein manifolds. Let
($X^k,g^k)$ be a Riemannian manifold, then its cone is $Y =\mathbb R_{\geq 0} \times X$ with metric
 \begin{equation}
   g^Y = \dd r^2 + r^2 g^k.
 \end{equation} 
 It has a singularity at $r=0$, unless $X$ is a round sphere, when the cone is simply flat Euclidean space.
 A simple calculation shows that $Y$ is Ricci-flat if and only if $X$ is Einstein, with normalization Ric$^k=
(k-1)\,g^k$. Furthermore, one can express the spin connection of $Y$ in terms of the connection on $X$, and finds
that it admits a parallel spinor if and only if $X$ has a so-called geometric Killing spinor $\epsilon$ \cite{Baer93},
satisfying
 \begin{equation}
   \Big( \nabla_a \pm \frac \ii2 \gamma_a \Big) \epsilon = 0.
 \end{equation} 
 Geometric Killing spinors induce a reduction of the structure group $K$ of $X$ to a proper subgroup of SO($k$),
in addition to the holonomy reduction of the cone to a group $G\subset \mbox{SO}(k+1)$. The following table gives the
classification of geometric Killing spinor manifolds and cones with reduced holonomy:
 \renewcommand{\arraystretch}{1.2}
  \begin{table}[H]\centering
 \begin{tabular}{cccc}  \toprule
  $\dim X$ & $X$ &  $K$ & $G$ \\ \hline  
  6&  nearly K\"ahler   & SU(3) & $G_2$\\
  7 &nearly parallel $G_2$ & $G_2$ & Spin(7)  \\
  $2m+1$ &Sasaki-Einstein  & SU($m$) & SU($m+1$)  \\
  $4m+3$ & 3-Sasakian & Sp$(m)$ & Sp$(m+1)$ \\
 $ m$ & $S^m$ & SO($m$) & $\{ 1\}$ \\
 \bottomrule
\end{tabular}
 \caption{Geometric Killing spinor manifolds with structure group $K$, and holonomy group $G$ on
the cone. Whereas even-dimensional round spheres do not admit a reduction of the structure group $K$,
except for $S^6$ which has a nearly K\"ahler structure, odd-dimensional spheres are always
Sasaki-Einstein  (also 3-Sasakian and nearly parallel $G_2$ where appropriate), and hence
admit a reduction to SU($m$).}\label{tab:Killingmfs} 
\end{table}
\noindent An important common property of geometric Killing spinor manifolds (except for even-dimensional spheres
$S^{2m},\, m\neq 3$) is the existence of
a canonical connection $\nabla^P$, with the following properties \cite{FI01,HN11}:
\begin{itemize}
 \item metric compatibility;
   \item holonomy group $K$; hence $\nabla^P$ admits a parallel spinor, which can be identified with the geometric
Killing spinor $\epsilon$, as well as a parallel 3-form $P$ and 4-form $Q^X$. It turns out that $Q^X\propto \dd P$;
 \item totally skew-symmetric torsion induced by the 3-form $P$;
  \item its curvature form $R^P$ satisfies the instanton equation $R^P\cdot \epsilon = 0$.
\end{itemize}
 Sasaki-Einstein and 3-Sasakian manifolds both admit a canonical 1-parameter family of metric deformations. The
connection $\nabla^P$ is compatible with this whole family of metrics, but its torsion form is only
skew-symmetric for a particular one, which differs from the original Einstein metric.

For $\dim X=3$ the 3-form $P$ coincides with the volume form and $Q^X=0$, whereas in all higher dimensions $Q^X\neq 0$.
This explains why the 3-dimensional case is somewhat special.

\subsection{Strings}
 According to our discussion in Section \ref{subsec:harmdef}, string-like supergravity solutions of the form
\eqref{strings:fields_general} can be obtained from
a harmonic function $h$ on a Ricci-flat manifold $Y$, which we now choose as a cone over an Einstein manifold $X^{k}$,
i.e. $Y= \mathbb R_{>0}\times X$. The simplest harmonic functions on $Y$ for $k\geq 2$ are of the form
\begin{equation}
  h(r) = 1 + \frac{ Q_e}{r^{k-1}},
\end{equation} 
with $Q_e$ a constant. String theory considerations imply that $Q_e \propto (\alpha')^{\frac {k-1}2} N$, with
integer $N$, to be identified with the number of strings. When $X = S^k$ and $Y= \mathbb R^{k+1}$, a more general
solution is 
 \begin{equation}
  h(\mathbf x) = 1 + \sum_a \frac{ Q_e^a}{|\mathbf x- \mathbf x_a|^{k-1}},
 \end{equation} 
 describing strings located at distinct points $\mathbf x_a \in \mathbb R^{k+1}$.

\subsection{Instantons}
 The tangent bundle of a cone with parallel spinors admits a 1-parameter family of instantons which resemble the
classical BPST instantons on $\mathbb R^4$. The construction was given in \cite{HN11}, based on earlier results for
$\mathbb R^7$ and $\mathbb R^8$ (the octonionic instantons) \cite{FN84,FN85,IP92,GN95} as well as certain
homogeneous
spaces \cite{ILPR09,HILP10,BILL10,HILP11}.

 The discussion is simplified by the observation that the cone metric is conformal to the cylinder metric $\dd \tau^2 +
g^k$, via a substitution $r= e^\tau$:
 \begin{equation}
   \dd r^2 + r^2 g^k = e^{2\tau } \big( \dd \tau^2 + g^k\big).
 \end{equation} 
Conformal invariance of the instanton equation implies that we can solve it on the cylinder $Z=\mathbb R\times
X$ instead of the cone. The canonical connection $\nabla^P$ pulls back from $X$ to $Z$, and the Ricci-flat
connection $\nabla^0$ of the cone can be considered as a connection on the cylinder as well. In the following I will
discuss the construction of instantons on cones of holonomy group $G= \mbox{SU}(2),\, G_2$ or Spin(7), and comment on
the case of SU($m$) and Sp($m$) later. Define the difference tensor as
 \begin{equation}
   A= \nabla^0 - \nabla^P\qquad \in \ \Gamma\big(T^* Z \otimes \mbox{End}(TZ)\big).
 \end{equation} 
An explicit expression for $A$ can be found in \cite{HN11}. Let $\psi(\tau)$ be a function of the cylinder variable
$\tau$, and define a connection
 \begin{equation}\label{gauge_field_ansatz}
   \nabla^\psi  = \nabla^P + \psi (\tau) A.
 \end{equation} 
Imposing the instanton equation on $\nabla^\psi$ leads to the equation 
\begin{equation}\label{InstEqPsi}
 \partial_\tau\psi  =2\psi(\psi-1) ,
\end{equation}
 which has two stationary solutions $\psi=0,1$, and a 1-parameter family of
interpolating solutions 
 \begin{equation}\label{InstSoltns}
   \psi = \Big(1 + e^{2(\tau-\tau_0)} \Big)^{-1}=\frac {\rho^2}{r^2+ \rho^2},
 \end{equation} 
 where $0< \rho=e^{\tau_0}< \infty$ is the parameter, and $r=e^\tau$. The full space of (bounded) solutions can be
parametrized by $0\leq \rho\leq \infty$, where $\rho=0$ and $\rho=\infty$ are the stationary solutions $\psi=0$ and
$\psi=1$. 
 While it appears that for $\psi=0$ and $\psi=1$ one obtains the canonical connection $\nabla^P$ and the Levi-Civita
$\nabla^0$ of the cone, it turns out that things are slightly more complicated due to the behaviour at the apex $r=0$.
 In the limit $\rho \rightarrow 0$ both the energy tr$|F^\psi|^2$ and the first Pontryagin class tr$(F^\psi\wedge
F^\psi)$
develop a delta function-like peak at $r=0$, so the connection $\nabla^{\psi=0}$ must be viewed as a point-like
instanton living at the apex. This interpretation is most compelling for the usual BPST instantons, obtained for
$X=S^3$, since then the field strength vanishes away from $r=0$ in the limit $\psi=0$. In higher dimensions there
always remains a finite energy and Pontryagin density at finite $r$, besides the singularity at $r=0$.
 
For $X=S^3$ one can also construct multi-instanton solutions whose centers can be arbitrary points in $\mathbb R^4$.
Interestingly, no one has succeeded so far in constructing multi-instantons on one of the higher dimensional cones, not
even octonionic multi-instantons have been found on $\mathbb R^7$ or $\mathbb R^8$. A generalization of the ansatz
\eqref{gauge_field_ansatz} to matrix valued functions $\psi(\tau)$ has been studied in \cite{IvanovaPopov12}.

 When $X$ is a Sasakian manifold, then there are two tensors $A^1$ and $A^2$ that one can add to the canonical
connection, and one has to allow for different $\tau$-dependence of their coefficients in order to solve the instanton
equation. While the qualitative behaviour of the solutions remains very similar to the one discussed above, the
differential equations involved become too difficult to be solved analytically in the Sasaki-Einstein case, and one
has to resort to numerical methods. Surprisingly, the 3-Sasakian case is simpler again and admits analytical solutions
similar to \eqref{InstSoltns}. See \S 4.2 and 4.3 in \cite{HN11}.

For $X=S^3$ the instantons \eqref{InstSoltns} reproduce the BPST instantons on $\mathbb R^4$, for $X=S^6$ with nearly
K\"ahler structure or $X=S^7$ with nearly parallel $G_2$-structure they give rise to the octonionic instantons on
$\mathbb R^7$ and $\mathbb R^8$, and in the 3-Sasakian case $X=S^{4m+3}$ one obtains a set of quaternionic instantons
found in \cite{CGK85,BIL08}. An equivalent ansatz for SU($m$)-instantons on cones over regular Sasaki-Einstein manifolds
had been proposed earlier by Correia \cite{Correia09,Correia10}.

\subsection{NS5-branes}
 As explained in Section \ref{subsec:confdef}, we can construct 5-brane solutions to heterotic supergravity by a
conformal deformation $g_0 \rightarrow e^{2f}g_0$ of the Ricci-flat cone metric. A convenient choice for the conformal
factor $f$ is to take it as a function of the radial variable $r$ only. Again, we will restrict attention to holonomy
groups $G= \mbox{SU}(2),\,G_2$ and Spin(7) for simplicity. The Casimir 4-form $Q$ on a cone assumes the
form 
 \begin{equation}
   Q = r^4 Q^X + r^3\dd r\wedge P,
 \end{equation}
 where $P$ and $Q^X$ are the canonical 3- and 4-forms on the base $X$. The 3-form $H$ will be proportional to
$P$, so that in particular $\dd H\neq 0$, unless $H$ itself vanishes, and in order to solve the Bianchi identity
\eqref{H-Bianchi} we need to choose non-trivial instanton connections $\tilde \nabla$ and $\nabla^A$. Setting
\begin{equation}
 \tilde
\nabla = \nabla^{\psi_1} \und \nabla^A=\nabla^{\psi_2},
\end{equation}  
 where both $\psi_1$ and $\psi_2$ satisfy the
instanton equation \eqref{InstEqPsi}, the Bianchi identity becomes a differential equation for $f$ \cite{HN11,GHLNP12}:
\begin{equation}
 \dot f e^{2f} = -\frac {\alpha'}{4r^2} \big( \psi_1 ^2-\psi_1\dot \psi_1 -\psi_2^2 +\psi_2\dot\psi_2\Big),
\end{equation} 
 where a dot denotes $\partial_\tau = r\partial _r$. The solution is given by
 \begin{equation}\label{f_sol}
   e^{2f} = \lambda^2 + \frac {\alpha'}{4r^2} \Big(\psi_1^2-\psi_2^2\Big),
 \end{equation} 
 where $\lambda$ is a volume modulus. In order for the solution to be well-defined in the full region
$0<r<\infty$ we impose that the scale parameters $\rho_1,\rho_2$ of $\psi_1$ and $\psi_2$ satisfy $\rho_1>\rho_2$. In
the generic case $0< \rho_2 <\rho_1 <\infty$ the full solution becomes
 \begin{equation}
  \begin{aligned}{}
    g& = \bigg(\lambda^2 + \frac{\alpha' } {4r^2} \bigg[\frac {\rho_1^4}{(\rho_1^2+ r^2)^2} - \frac
{\rho_2^4}{(\rho_2^2+ r^2)^2}\bigg] \bigg) \big(\dd r^2 + r^2 g^k \big)\, , \\
   H&= \frac {\alpha'}{2\beta}
\bigg(\frac{\rho_1^6+3\rho_1^4r^2}{(\rho_1^2+r^2)^3} -\frac{\rho_2^6+3\rho_2^4r^2}{(\rho_2^2+r^2)^3} \bigg)\, P\, , \\
   e^{2(\phi-\phi_0)} &= \bigg(\lambda^2 + \frac{\alpha' } {4r^2} \bigg[\frac {\rho_1^4}{(\rho_1^2+ r^2)^2} - \frac
{\rho_2^4}{(\rho_2^2+ r^2)^2}\bigg] \bigg)^{{k/3}}.
  \end{aligned}
 \end{equation} 
 We can view the contributions of $\nabla^A$ and $\tilde \nabla$ as 5-branes and anti 5-branes, respectively. To get a
non-vanishing total 5-brane number we set $\rho_1 =\infty$ and hence eliminate the anti 5-brane. The resulting gauge
5-brane for $X=S^3$ coincides with Strominger's solution \cite{Strom90}, for $X=S^6$ with one found by Ivanova and
G\"unaydin-Nicolai \cite{Iva93,GN95}, and for $X=S^7$ with the Harvey-Strominger octonionic superstring soliton
\cite{HS90}. For $X=S^3$ it is also possible to construct multi-brane solutions, based on multi-instantons. 
 Let us then consider the limit of a (partially) localized 5-brane, i.e. $(\rho_1,\rho_2) \rightarrow
(\infty,0)$:
\begin{equation}\label{NS5-brane_sol_cone}
 \begin{aligned}{}
   g&= \Big(\lambda^2 + \frac{\alpha'}{4r^2} \Big) \big(\dd r^2 + r^2 g^k\big) ,\\
   H &= \frac {\alpha'}{2\beta}\, P,\\
   e^{2(\phi-\phi_0)} &= \Big(\lambda^2 + \frac{\alpha'}{4r^2} \Big)^{k/3}.
 \end{aligned}
\end{equation} 
 In the near horizon limit $r\rightarrow 0$ this shows the linear dilaton behaviour characteristic of NS5-branes:
 \begin{equation}
  \begin{aligned}{}
    g &= \frac{\alpha'}4 \big(\dd \tau^2 + g^k\big) ,\\
    H &= \frac {\alpha'}{2\beta} \, P,\\
    \phi &= -\sfrac k3  \,\tau +\mbox{const},
  \end{aligned}
 \end{equation} 
 in terms of the cylinder variable $\tau = \log(r)$. The full solution \eqref{NS5-brane_sol_cone} describes a single
NS5-brane on the Ricci-flat cone $Y= \mathbb R_{>0}\times X$, and is a generalization of the classical 5-brane solution
of \cite{CHS91a}, which has $X=S^3$. 

When the cone has  holonomy group SU($m$) or Sp($m-1)$ with $m>2$, the construction of instantons and supergravity
solutions gets modified, but works in essentially the same way. As explained before, for SU($m$) only numerical
solutions can be obtained, whereas Sp$(m$) admits full analytical solutions. The qualitative behaviour is always the
same as for the solutions above.

\subsection{Fractional strings}\label{ssec:fracstring}
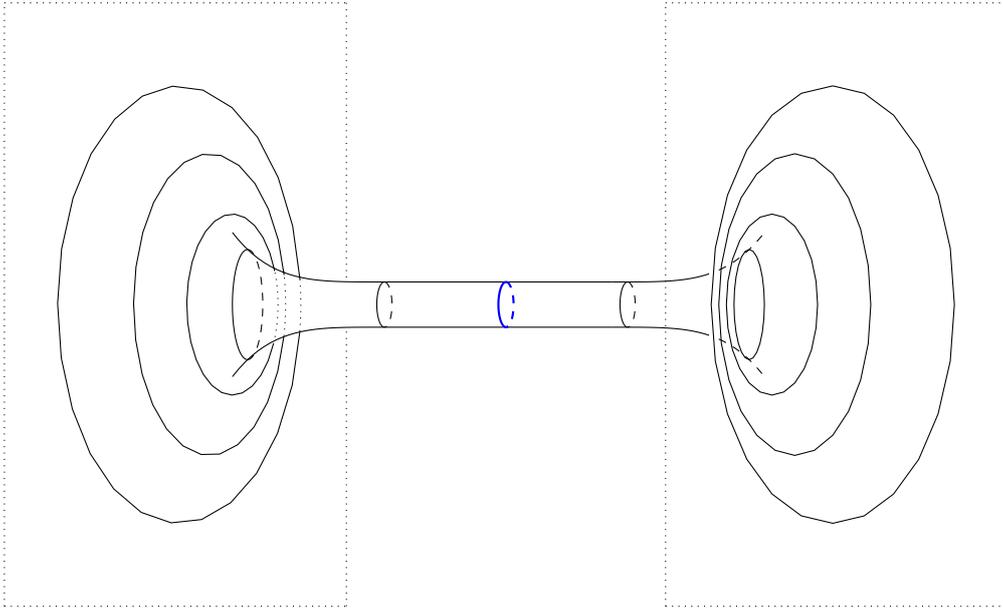
\begin{figure}[H]
\begin{center}
\begin{tikzpicture}[domain=0:2]  
 \draw[dashed] plot ({0.4+\x/10*3},{0.3 + (0.45*( 1.4+\x/10*3))^4});            % tube rechts oben
 \draw[dashed] plot ({0.4+\x/10*3},{-0.3 - (0.45*(1.4+\x/10*3))^4});            % tube rechts unten
  \draw plot ({\x/11*7-1},{0.3 + (0.45*( \x/11*7))^4});            % tube rechts oben
  \draw plot ({\x/11*7-1},{-0.3 - (0.45*(\x/11*7))^4});            % tube rechts unten
 \draw plot ({-4-\x},{0.3 + (0.45*( \x))^4});            % tube links oben
 \draw plot ({-4-\x},{-0.3 - (0.45*( \x))^4});            % tube links unten
 \draw[color=black] plot ({0.8 - 0.2*sin((\x r)*pi)},{(0.3+(0.45*1.8)^4)*cos((\x r)*pi)}); % rechts: Kreis innen
 \draw[color=black] plot ({1.1 - 0.6*sin((\x r)*pi)},{1.2*cos((\x r)*pi)}); % Kreis 2
 \draw[color=black] plot ({1.4 - 1*sin((\x r)*pi)},{2*cos((\x r)*pi)}); % Kreis 3
 \draw[color=black] plot ({1.9 - 1.6*sin((\x r)*pi)},{2.9*cos((\x r)*pi)}); % Kreis 4
 \draw plot({-5.8 - 0.2*sin((\x r)*pi/2)},{(0.3+(0.45*1.8)^4)*cos((\x r)*pi/2)}); % links: Kreis innen
 \draw[dashed] plot({-5.8 + 0.2*sin((\x r)*pi/2)},{(0.3+(0.45*1.8)^4)*cos((\x r)*pi/2)}); % links: Kreis innen
 \draw[dotted] plot({-6 - 0.6*sin((\x r)*pi/8-111)},{1.2*cos((\x r)*pi/8-111)}); % Kreis 2
 \draw plot({-6 - 0.6*sin(((\x r)-25)*pi/1.15)},{1.2*cos(((\x r)-25 )*pi/1.15)}); % Kreis 2
 \draw[dotted] plot({-6.3 - 1*sin((\x r)*pi/14-102)},{2*cos((\x r)*pi/14-102)}); % Kreis 3
 \draw plot({-6.3 - 1*sin(((\x r)-26.8)*pi/1.07)},{2*cos(((\x r)-26.8 )*pi/1.07)}); % Kreis 3
 \draw[dotted] plot({-6.7 - 1.6*sin((\x r)*pi/25 - 97)},{2.9*cos((\x r)*pi/25 -97)}); % Kreis 4
 \draw plot({-6.7 - 1.6*sin(((\x r)-27.6)*pi/1.04)},{2.9*cos(((\x r)-27.6 )*pi/1.04)}); % Kreis 4
 \draw[dotted] (-0.3,-4) -- (4.2,-4); \draw[dotted] (-0.3,-4) -- (-0.3,4); \draw[dotted] (-0.3,4) -- (4.2,4);
\draw[dotted] (4.2,4) -- (4.2,-4); % right rectangle
 \draw[dotted] (-9,-4) -- (-4.5,-4); \draw[dotted] (-9,-4) -- (-9,4); \draw[dotted] (-9,4) -- (-4.5,4); \draw[dotted]
(-4.5,4) -- (-4.5,0.3); \draw[dotted] (-4.5,-4) -- (-4.5,-0.3); % left rectangle
 \draw(-4,0.3)--(-1,0.3); \draw(-4,-0.3) -- (-1,-0.3); % tube
 \draw plot({-4 - 0.1*sin((\x r)*pi/2)},{0.3*cos((\x r)*pi/2)}); % inner circle 1
 \draw[dashed] plot({-4 + 0.1*sin((\x r)*pi/2)},{0.3*cos((\x r)*pi/2)}); % inner circle 1
 \draw plot({-0.8- 0.1*sin((\x r)*pi/2)},{0.3*cos((\x r)*pi/2)}); % inner circle 1
 \draw[dashed] plot({ -0.8+ 0.1*sin((\x r)*pi/2)},{0.3*cos((\x r)*pi/2)}); % inner circle 1
 \draw[color=blue,thick] plot({-2.4 - 0.1*sin((\x r)*pi/2)},{0.3*cos((\x r)*pi/2)}); % inner circle 1
 \draw[color=blue,dashed,thick] plot({-2.4 + 0.1*sin((\x r)*pi/2)},{0.3*cos((\x r)*pi/2)}); % inner circle 1
\end{tikzpicture}\caption{Potential wormhole geometry of a modified fractional string solution, obtained by removing
the strings and 5-branes, whose original location is the central blue circle, and assuming reflection symmetry in the
radial variable. The picture is somewhat simplifying,
since two non-trivial dimensions are suppressed. The asymptotic solution for very small or large values of the radial
coordinate $r$ is a
direct product of the Ricci-flat cone $\mathbb R_{>0} \times X^k$ with $\mathbb R^{1,1}$, whereas a
neighbourhood of the circle approximates AdS$_3\times X^k$. Additionally, there is a flat torus filling up the
missing dimensions.}\label{fig:wormhole}
\end{center}
\end{figure}

 Having constructed both string and 5-brane solutions of heterotic supergravity on cones, one can go ahead and look for
superpositions of both \cite{GHLNP12}. In the simplest case of a a cone over $S^3$ such solutions have been found
before, see for instance \cite{LLOP99}, and there it turns out that the structure of the distinct ingredients remains
completely unchanged in the full solution. This is not the general rule in higher dimensions, however; whereas the
5-brane solution remains intact, the
harmonic function defining the string gets modified. We consider again the case of holonomy groups SU(2), $G_2$ or
Spin(7) on the cone, although similar solutions can be obtained for SU($m$) and Sp($m$).
Let ($X^{k},g^k$) be a geometric Killing spinor manifold of
dimension $k\leq 7$, with nearly K\"ahler or nearly parallel $G_2$-structure, or $X=S^3$. Consider the spacetime
 \begin{equation}
  \mathbb R^{1,1}\times \mathbb R_{>0} \times X^k \times \mathbb T^{7-k},
 \end{equation} 
with metric
 \begin{equation}\label{String-5brane:metric}
  \begin{aligned}{}
     g &= h^{-1} (-\dd t^2+ \dd x^2) + e^{2f}\big(\dd r^2  + r^2 g^k\big) + g_{\mathbb T^{7-k}}, \\
  \end{aligned}
 \end{equation} 
 where both $h$ and $f$ are functions of the radial variable $r$. We can solve the BPS equations \eqref{BPSeqtns} by
superposing the string and 5-brane ans\"atze for $H$ and $\phi$ from \eqref{strings:fields_general} and
\eqref{5branes:fields_general}: 
 \begin{equation}\label{String-5brane:flux}
  \begin{aligned}{}
    H &= \dd h^{-1} \wedge \dd t\wedge \dd x -{\beta}^{-1} r^3\partial_r e^{2f} \, P,\\
    e^{2(\phi-\phi_0)} &= h^{-1} e^{2k f/3}.
  \end{aligned}
 \end{equation} 
 As before, the gauge fields $\tilde \nabla$ and $\nabla^A$ can be chosen as instantons on the cylinder over $X$. Then
the Bianchi identity \eqref{H-Bianchi} and field equation \eqref{H-eqtn} for $H$ determine the conformal and warp
factors $f$ and $h$. The Bianchi identity remains unchanged, so $f$ will be of the form
\eqref{f_sol}, whereas the field equation couples to $f$ and gets modified:
 \begin{equation}\label{fracstringheq}
   \partial_\tau \Big[\dot h\, \exp\Big\{\sfrac13 \big(k-3 \big) f +\big(k-1 \big)\tau\Big\} \Big] =0.
 \end{equation}  
 Again the dot denotes a derivative with respect to $\tau = \log(r)$. Note that this equation differs from the
condition for $h$ to be a harmonic function on the conformally rescaled geometry $e^{2f} (\dd r^2 + r^2 g^k)$, which
reads
\begin{equation}\label{fracstringharmoniceq}
   \partial_\tau \Big[ \dot h \, \exp\Big\{\big(k-1 \big)( f+\tau)\Big\}\Big] =0.
 \end{equation} 
An important aspect of \eqref{fracstringheq} is that for $X=S^3$ the $f$-dependent term drops out, and $h$ remains a
harmonic function on $\mathbb R ^4$. This would not be the case for \eqref{fracstringharmoniceq}.
In higher dimensions, to solve \eqref{fracstringheq} for $h$, one has to plug in the
result for $f$ from the Bianchi identity. In the NS5-brane limit $(\rho_1,\rho_2)\rightarrow
(\infty,0)$ this can be done explicitly (except again for Sasaki-Einstein manifolds $X$). The results can be found
in \S 5.5 of \cite{GHLNP12}; they look rather complicated, but have a simple limiting behaviour for $r\rightarrow
0$ and
$r\rightarrow \infty$. In the latter case one gets back the Ricci-flat vacuum solution, whereas the near horizon limit
$r\rightarrow 0$ has an AdS$_3$ geometry: here 
 \begin{equation}
   h^{-1} \propto r^{ 2k/3 },\qquad e^{2f} = \frac {\alpha'}{4r^2},
 \end{equation} 
 and upon substituting $s = h^{-1/2} \propto r^{k/3}$ we arrive at the following solution:
%
%\begin{equation}
% \begin{aligned}{}
% g &= s^2(-\dd t^2 +\dd x^2) +{\alpha'} \Big(\frac{\beta}{2\kappa}\Big)^2  \frac{\dd s^2}{s^2} +\frac {\alpha'}4 g^X
%+ g_{\mathbb T^{7-k}} ,\\ 
%  H &= \dd(s^2) \wedge \dd t\wedge \dd x + \frac {\alpha'}{2\beta} P ,\\
%  \phi &= \mbox{const}.
% \end{aligned}
%\end{equation}
%  
\begin{equation}\label{String-5brane-AdS}
 \begin{aligned}{}
 g &= s^2(-\dd t^2 +\dd x^2) + \frac {9\alpha'}{4k^2}  \frac{\dd s^2}{s^2} +\frac {\alpha'}4 g^k
+ g_{\mathbb T^{7-k}} ,\\ 
  H &= \dd(s^2) \wedge \dd t\wedge \dd x + \frac {\alpha'}{2\beta} P ,\\
  \phi &= \mbox{const}.
 \end{aligned}
\end{equation}
 The coordinates $(t,x,s)$ parametrize a patch of AdS$_3$, and the metric is a direct product metric on
AdS$_3\times X^{k}\times \mathbb T^{7-k}$. For the special case of $X=S^3$ the solution coincides with the D1-D5-brane
system, except that the RR-flux gets replaced by
NS-flux. Note that the locus $\{s=0\}$ inside AdS$_3$ is a regular submanifold, so it looks as if the
previously singular tip of the cone has been blown up by the strings and 5-branes. However, the solution
\eqref{String-5brane-AdS} is valid only for $s>0$, and all fields acquire additional delta-function source terms, which
can be derived by considering the solution with $(\rho_1,\rho_2) = (\infty,\rho)$ for finite $\rho$, and
taking the small instanton limit $\rho \rightarrow 0$. Therefore the fractional string remains singular
at $s=0$, in accordance with the observation that the Pontryagin density of the gauge field concentrates at $s=0$ in
this limit.
\begin{figure}[H]
\centerline{
        \begin{overpic}[scale=0.5]%
            {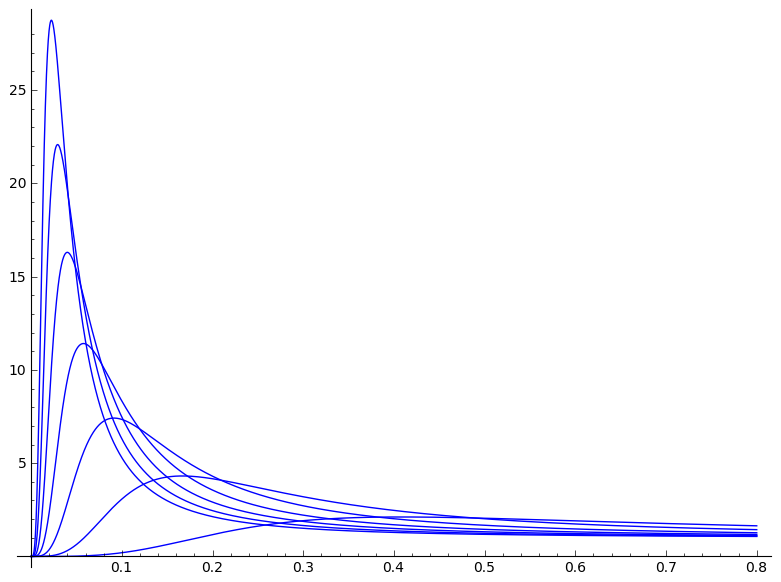} \put(285,5){\large $r$}\put(-19,200){\large $F_\rho(r)$}
      \end{overpic}}
\caption{Plot of the coefficient $F_\rho(r) = \frac {3\rho^2 r^4 +r^6}{(\rho^2+r^2)^3}$, defined by $H=
\frac{\alpha'}{2\beta} F_\rho(r) P$ for a supergravity gauge 5-brane with ($\rho_1,\rho_2)=(\infty,\rho)$. The  
scale parameters in the plot are $\rho = 1/n^2,$ $n=2,\dots,8$. In the limit $\rho \rightarrow 0$ a delta peak at $r=0$
arises. The other supergravity fields contain similar delta source terms.}
\end{figure}
%
%\begin{figure}[H]
%\begin{center}
%\includegraphics[width=0.7\textwidth, clip]{sage0.png}
%\caption{Plot of the coefficient $F_\rho(r) = \frac {3\rho^2 r^4 +r^6}{(\rho^2+r^2)^3}$, defined by $H=
%\frac{\alpha'}{2\beta} F_\rho(r) P$ for a supergravity gauge 5-brane, with scale parameters $\rho = 1/n^2,$
%$n=2,\dots,8$. In the limit $\rho \rightarrow 0$ a delta peak at $r=0$ arises.}
%\label{fig:se5sol}
%\end{center}
%\end{figure}
%
Due to the persistence of the singularity at $r=s=0$ it is not possible to determine directly the world-volume of the
5-brane in the fractional string solution. But here, contrary to a pure gauge brane, the singularity can be removed, by
simply setting the value of all fields at $r=0$ equal to their limiting value as $r \rightarrow 0$. Then, as explained
above, the topology of the hypersurface $\{r=0\}$ will be that of a submanifold of AdS$_3$, times $X^k\times \mathbb
T^{7-k}$. The resulting solution can be extended to negative values of $r$ as well, and since the gauge field of the
NS5-brane coincides with the pull-back of the canonical connection on $X$ for all $r>0$, it
appears to be a reasonable assumption that this will be the case globally for the modified solution. There is a
 coordinate singularity at $r=0$, as one can see from the metric in \eqref{String-5brane-AdS}, so in order to
determine the continuation of the supergravity fields to negative $r$-values one needs to find a better set of
coordinates. Similarly to AdS$_3$, one may suspect that $r$ will be globally defined and take arbitrary values in
$\mathbb R$, whereas $t$ and $x$ need to be replaced around $r=0$. Possibly, the geometry of the new solution will look
like in Figure \ref{fig:wormhole}, where reflection symmetry in $r$ is assumed.
From the gauge-theory point of view the existence of the modified supergravity solution may be motivated as follows. One
expects that every instanton gauge field can be lifted to a heterotic supergravity solution. For a certain
class of instantons on the cone we have constructed such a lift, which is not unique in this case but can for instance
include strings. But there is another instanton, namely the canonical connection $\nabla^P$ on $X$, which
essentially coincides with the small instanton on the cone away from $r=0$. The modified solution is exactly the
supergravity lift of this connection. There is some evidence that $\nabla^P$ always forms an isolated point in
instanton moduli space, which means that the corresponding supergravity solution should not admit deformations either
\cite{Xu09,CH12}.

In the simplest example $X=S^3$ the full metric for $r>0$ reads
\begin{equation}
   g =  \Big(1 + \frac {Q_e}{r^2} \Big)^{-1}\big(-\dd t^2+\dd x^2) + \Big(1 + \frac {Q_m}{r^2} \Big) \big(\dd r^2 +r^2
g_{S^3}\big), 
\end{equation} 
with positive constants $Q_e,Q_m$, and the task is to find a continuation through $r=0$. The $H$-field and dilaton are
\begin{equation}
 \begin{aligned}{}
    H&= \frac {2rQ_e}{(r^2+Q_e)^2} \dd r\wedge \dd t\wedge \dd x + 2Q_m\mbox{Vol}_{S^3} , \\
    e^{2(\phi-\phi_0)} &=   \frac {r^2 + Q_m}{r^2 + Q_e}.
 \end{aligned}
 \end{equation} 
 Assuming the $r<0$ solution to be of the same form, one concludes that the dilaton is smooth at $r=0$ and both metric
and $H$-field are at least continuous.

Now suppose the conjectured completely regular solution to be known. Then it must be possible to extend also the
instantons \eqref{gauge_field_ansatz} to negative $r$-values. One can then envisage the determination of the singular
support of the gauge field in the small instanton limit, which will be contained in the region $r\leq 0$, but
presumably even in the 9-dimensional hypersurface $\{r=0\}$. Although we cannot currently prove it, it appears very
likely that the dimension of the singular support will now assume its maximal value six, and wrap a smeared
$(k-3)$-cycle $\Sigma$ in $X$, in accordance with the 5-brane interpretation.

To summarize, the proposed method to determine the world-volume of an instanton based NS5-brane is as follows. Suppose
a family of instanton gauge fields on a Ricci-flat manifold $Y$ to be given, depending on a scale parameter $\rho$. The
singular support of the small instanton ($\rho=0$) is a submanifold of codimension at least four, which
we want to
identify with the brane. It can happen, however, that the codimension is larger than four, in which case we need to
identify a more suitable background $\tilde Y$ which admits the given gauge field as an instanton as well. Determine the
supergravity solutions for the family of instantons on $Y$ for finite $\rho$, which correspond to smeared 5-branes.
Add background strings, take the small instanton limit and eliminate the singularity at the supposed string and brane
world-volume to obtain a completely smooth supergravity solution. Then extend the original smooth instantons to this new
solution $\tilde Y$, which is a Lorentzian manifold of two dimensions higher than $Y$, due to the
background strings. Take again the small-instanton
limit, and determine its singular support, which now hopefully turns out to be of codimension four and wrap some
calibrated
submanifolds in $\tilde Y$. 
This would also fit nicely into the picture of D-branes wrapping calibrated submanifolds in Calabi-Yau manifolds
 \cite{Gaunt03,Aspinwall04}. In the case $X=S^3$ the cycle $\Sigma$ can presumably be viewed as the weighted sum of all
points in $S^3$, so that the 5-brane is evenly smeared over $S^3$. The reinterpretation is not strictly necessary in
this case, however, because the singular support of a small BPST instanton on $\mathbb R^{9,1}$ is already
6-dimensional; replacing the origin of $\mathbb R^4\subset \mathbb R^{9,1}$ by a copy of $S^3$ only amounts to smearing
the world-volume.

The AdS$_3$-limit of the supergravity solution suggests a holographic duality to a 2-dimensional conformal field
theory. The asymptotic isometry algebra of the backgrounds has been determined in \cite{GHLNP12}, and was found to be
of the form 
\begin{equation}\label{sisom_generalform}
  \mathfrak g_0  \oplus \mathfrak g
\end{equation} 
 in all cases, where $\mathfrak g$ is some super Lie algebra with
bosonic subalgebra $\mathfrak g_0$, see Table \ref{tab:sisoms}. This form of the algebra suggests an interpretation in
terms of heterotic sigma models; since these have purely bosonic right-moving modes and superpartners for the
left-moving modes, their symmetry algebra is generically of the form \eqref{sisom_generalform}, and they provide natural
candidates for holographically dual field theories. Such an interpretation has been advocated for similar heterotic
supergravity solutions of the form AdS$_3\times S^2 \times \mathbb T^5$ in a couple of papers
\cite{DabhMurthy07,LSS07,KLS07,Duff08}. Clearly, this question deserves further study.
 \renewcommand{\arraystretch}{1.2}
  \begin{table}[H]\centering
 \begin{tabular}{c|c|c|c}\toprule
   $\dim X$ & $X$ & $\mathfrak g_0$ &  $\mathfrak{g}$ \\ \hline 
 7 &  nearly parallel $G_2$&  $\mathfrak{so}(2,1) $ & $\mathfrak{osp}(1|2)$  \\
5, 7 & Sasaki-Einstein & $\mathfrak{so}(2,1)\oplus \mathfrak {u}(1)$
&$\mathfrak{osp}(2|2)$ \\
  7 & 3-Sasakian &  $\mathfrak{so}(2,1)\oplus \mathfrak {sp}(1)$  &  $\mathfrak{osp}(3|2)$ 
\\ 
  6 & nearly K\"ahler & $\mathfrak{so}(2,1)$  & $\mathfrak{osp}(1|2)$ \\
3 & Sasaki-Einstein & $\mathfrak{so}(2,1) \oplus \mathfrak{su}(2)$ & $\mathfrak{psu}(1,1|2)$
\\ \bottomrule
\end{tabular}
 \caption{Super isometry algebras $\mathfrak g$ of the fractional strings.}\label{tab:sisoms}
\end{table}
 The fractional string near horizon AdS$_3$-solutions to the gravitino and dilatino equations had been found
previously in \cite{KO09},
without considering the Bianchi identity, however.

\section*{Acknowledgements}
I would like to thank Derek Harland for useful comments.
This work was supported by the German Science Foundation (DFG) under the Collaborative Research Center (SFB) 676
``Particles, Strings and the Early Universe''.

\end{document}